\documentclass[aps,prl,twocolumn,superscriptaddress,showpacs]{revtex4}

\pdfoutput=1

\usepackage{amsmath,graphicx,amssymb,braket,xcolor,subfigure,upgreek,bbold}
\usepackage[colorlinks, linkcolor=blue, citecolor=blue, urlcolor=blue, breaklinks=true]{hyperref}

\begin{document}

\title{Adiabatic cooling of bosons in lattices to magnetic ordering}

\author{Johannes Schachenmayer}
\affiliation{JILA, NIST, Department of Physics, University of Colorado, 440 UCB, Boulder, CO 80309, USA}
\author{David M. Weld}
\affiliation{Department of Physics and California Institute for Quantum Emulation, University of California, Santa Barbara, CA 93106, USA}
\affiliation{MIT-Harvard Center for Ultracold Atoms, Research Laboratory of Electronics, Department of Physics, Massachusetts Institute of Technology, Cambridge, Massachusetts 02139, USA}
\author{Hirokazu Miyake}
\author{Georgios A. Siviloglou}
\affiliation{MIT-Harvard Center for Ultracold Atoms, Research Laboratory of Electronics, Department of Physics, Massachusetts Institute of Technology, Cambridge, Massachusetts 02139, USA}
\author{Andrew J. Daley}
\affiliation{Department of Physics and SUPA, University of Strathclyde, Glasgow G4 0NG, Scotland, UK}
\affiliation{Department of Physics and Astronomy, University of Pittsburgh, Pittsburgh, Pennsylvania 15260, USA}
\author{Wolfgang Ketterle}
\affiliation{MIT-Harvard Center for Ultracold Atoms, Research Laboratory of Electronics, Department of Physics, Massachusetts Institute of Technology, Cambridge, Massachusetts 02139, USA}

\date{March 25, 2015}

\pacs{37.10.Jk, 67.85.Hj, 42.50.-p, 03.65.Yz}

\begin{abstract}
We suggest and analyze a new scheme to adiabatically cool bosonic atoms to picokelvin temperatures which should allow the observation of magnetic ordering via superexchange in optical lattices. The starting point is a gapped phase called the spin Mott phase where each site is occupied by one spin-up and one spin-down atom.  An adiabatic ramp leads to an $xy$-ferromagnetic phase. We show that the combination of time-dependent density matrix renormalization group methods with quantum trajectories can be used to fully address possible experimental limitations due to decoherence, and demonstrate that the magnetic correlations are robust for experimentally realizable ramp speeds. Using a microscopic master equation treatment of light scattering in the many-particle system, we test the robustness of adiabatic state preparation against decoherence. Due to different ground-state symmetries, we also find a metastable state with $xy$-ferromagnetic order if the ramp crosses to regimes where the ground state is a $z$-ferromagnet.  The bosonic spin Mott phase as the initial gapped state for adiabatic cooling has many features in common with a fermionic band insulator, but the use of bosons should enable experiments with substantially lower initial entropies.
\end{abstract}

\maketitle

A major goal in the field of ultracold atoms is to reach picokelvin temperatures in optical lattices and observe new spin-ordered quantum phases \cite{McKay2011,Bloch2012}.  Such low temperatures are necessary due to the smallness of superexchange (second order tunneling) matrix elements \cite{Lewenstein2012} which determine the transition temperature to magnetically ordered phases \cite{Capogrosso-Sansone2010,Ho2007,Jordens2010}. The current strategy is to cool atoms by evaporative cooling, and then continue with some form of adiabatic cooling.  Adiabatic processes can dramatically lower the temperature of a system, if external parameters are slowly varied with respect to the level spacing between excited states of the system \cite{Ho2007,Medley2011,Rabl2003,Trebst2006,Kantian2010,Lubasch2011, Sorensen2010,Gammelmark2013}.  Since adiabatic processes conserve entropy, one should select an initial state which can be prepared with very low entropy.
\begin{figure}[t]
  \centering
\includegraphics[width=\columnwidth]{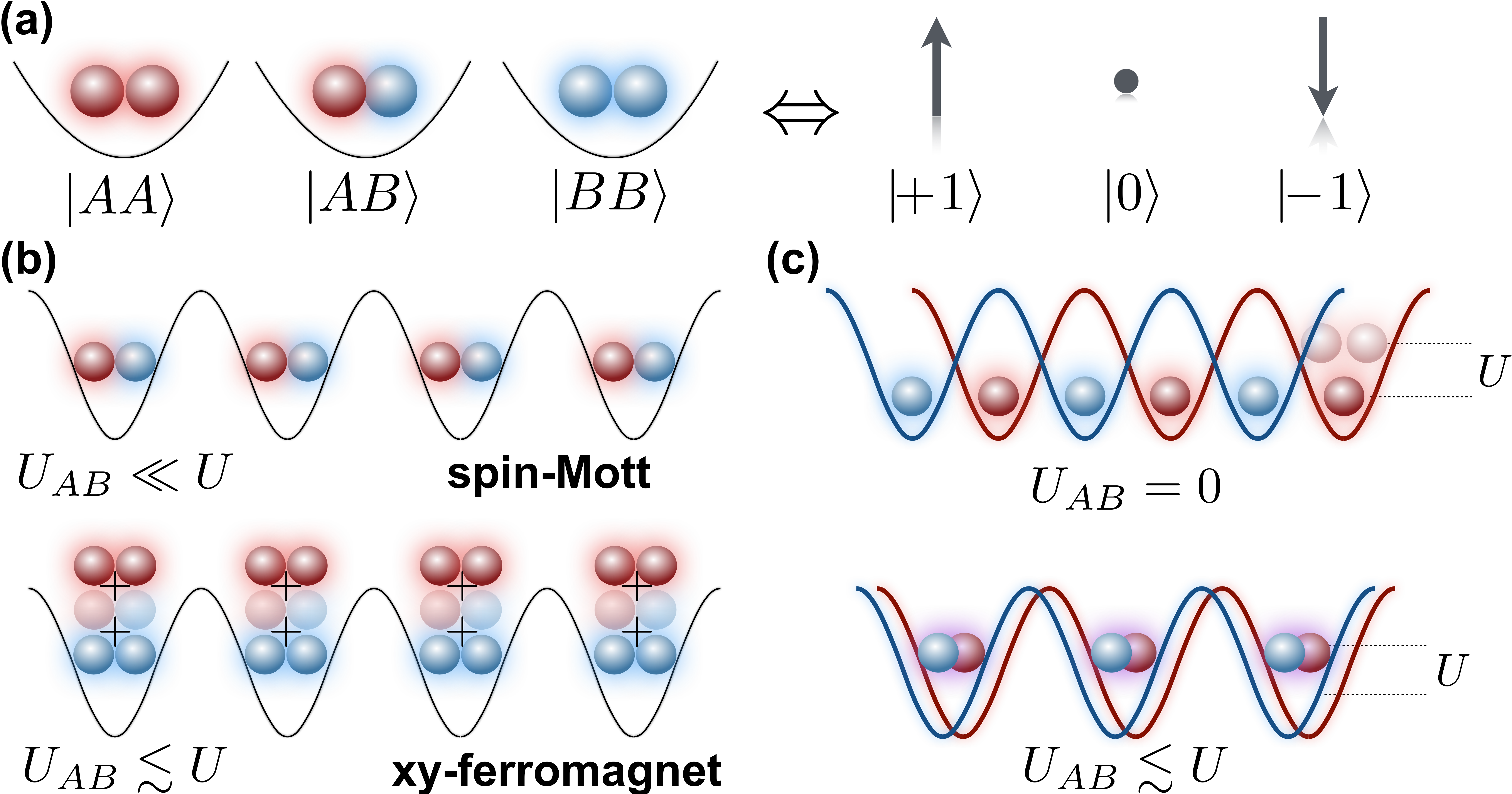}
\caption{{\it Setup for adiabatic preparation of magnetic states --} (a) Two-component bosons on a single lattice site with occupation number two and strong interactions can be represented as three different spin-1 states. (b) When the inter-component interaction $U_{AB}$ is negligible compared to the intra-component interaction $U$, the ground state of the system corresponds to a spin Mott state, for $U_{AB}\lesssim U$ to a planar $xy$-ferromagnetic state. (c) Spin-dependent lattices can be used to adibatically tune the system from a spin Mott state to an $xy$-ferromagnetic regime.  \label{fig:1}}
\end{figure}

The use of adiabatic ramps starting from a band insulator of fermionic atoms has been proposed for production of a variety of states \cite{Rabl2003,Trebst2006,Kantian2010,Lubasch2011}. These involve a ramp from states with a large gap that can be prepared with low entropy to a state with a much smaller gap, and often spin-ordering, generally making use of a superlattice potential to delocalize the atoms and select filling factors. For realisation of ordered states, bosonic atoms could provide significant advantages because evaporative cooling allows for the realisation of much lower entropies for bosons than for fermions \cite{McKay2011}. However, it has been difficult to find an equivalent of the band insulator state that can be straight-forwardly realised in an experiment. Here, we show that the spin Mott state in a two-component bosonic system \cite{Altman2003,Kuklov2003,Duan2003,Powell2009,Hubener2009,Capogrosso-Sansone2010} can play the role of the band insulator for the fermionic system, and that it can be prepared with low entropy from two independent Mott insulators in spin-dependent lattices \cite{Jaksch1999,Mandel2003,mckay2010,Gadway2010}. Using the control offered by such lattices, we can vary the inter-component interactions, and produce a ramp into a state of $xy$-ferromagnetism, driven by a spin-exchange term \cite{Altman2003,Kuklov2003,Duan2003} (see Fig.~1). Using time-dependent density matrix renormalization group techniques (t-DMRG) \cite{White2004,Daley2004,Verstraete2008,Schollwock2011} we show that this produces a state with high fidelity for realistic timescales in the experiment. A key question in all adiabatic preparation schemes is whether they can be robust in the presence of noise and dissipation. Due to the near-resonant nature of the spin-dependent lattice, light scattering is the limiting factor in this scheme \cite{Gordon1980,Dalibard1985,Gerbier2010,Pichler2010}. We compute the dynamics incorporating a microscopic treatment of the corresponding decoherence, and show that the magnetic order is surprisingly robust. This paves the way towards realisation of quantum magnetic order with ultracold atoms in an optical lattice. 

\textit{Low entropy bosons on a lattice -- } Bosons have advantages for reaching very low temperatures since the entropy $S/N$ per particle $S/N k_B= 3.6 (T/T_c)^3$ drops rapidly for temperatures $T$ below the BEC transition temperature $T_c$, and for almost pure condensates becomes almost unmeasurably small, of order 0.05. Magnetic ordering typically requires entropies below $\ln(2)=0.69$.  In contrast, for fermions, the entropy below the Fermi temperature $T_F$ is linear in temperature, $S/N k_B= {\pi}^2  (T/T_F)$ and values of $0.5$ are typically reached at $T/T_F=0.05$. Loading atoms into an optical lattice reduces the temperature (since this increases the effective mass), but leaves the total entropy constant.  However, if a gapped phase is formed in the center of a harmonic trapping potential -- a band insulator for fermions or Mott insulator for bosons -- then the entropy will accumulate at the edge of the cloud. Single-site imaging showed that Mott shells with one atom per site can have less than 1\% defects, with local entropies below $S/(N k_B ) <0.1$ \cite{Simon2011,Sherson2010}.  The challenge is now to realize such low entropies with a `spinful' system which has the spin degree of freedom and suitable interactions for magnetic ordering.

\textit{Adiabatic cooling --} Recently, we addressed this problem by introducing spin gradient demagnetization cooling of ultracold atoms \cite{Medley2011}. Two bosonic systems (spin-up and spin-down) were prepared in the Mott insulating phase, but separated by a strong magnetic field gradient.  Reducing the gradient mixes the two spins and reduces the temperature since kinetic entropy is transferred to spin entropy. However, beyond the proof-of-principle demonstration, this scheme has the major drawback that a macroscopic transport of atoms through the cloud is needed for the spin mixing.  This issue has a very elegant solution for fermions, where one can prepare a band insulator and, by doubling the period of the lattice using superlattices, adiabatically connect to an anti-ferromagnetic phase at half filling (for each spin component) \cite{Rabl2003,Trebst2006,Kantian2010,Lubasch2011}.  For fermions, another form of adiabatic cooling has been recently realized by ramping a lattice from isotropic to anisotropic tunneling \cite{Greif2013}, effectively cooling magnetic correlations in one direction by transferring  entropy to the other spatial direction.

Here, we address the major missing piece for bosons, how to adiabatically connect the low entropy Mott phase to a magnetically ordered phase. The basic idea is to combine spin gradient demagnetization cooling with spin-dependent lattices \cite{Jaksch1999,Mandel2003,mckay2010,Gadway2010}.  Spin-dependent lattices can be regarded as a (fictitious) alternating magnetic field gradient \cite{Sorensen2010}, separating spin-up and spin-down on each site, as shown in Fig.~1b.  In such lattices, it is possible to prepare two non-interacting Mott phases (for spin-up and spin-down).  The spin-up atoms reside on interstitial sites with respect to the spin-down lattice.  By ramping down the spin-dependent lattice we can fully mix the two Mott insulators.  This requires only microscopic motion of the atoms (by less than one lattice constant), in contrast to the previously demonstrated spin gradient demagnetization cooling.

\textit{Model and sketch of ground states --} This simple concept can be realised in a two-component Bose-Hubbard model.  Within the lowest Bloch band of the lattice, two-component bosons denoted $A$ and $B$ are well described by the two-component Bose-Hubbard Hamiltonian ($\hbar \equiv1$), 
\begin{align}
  \mathcal{H}
  =
  &- J \sum_{\langle j,l\rangle } \left( \hat a^\dag_{j} \hat a_{l} + \hat b^\dag_{j} \hat b_{l}  \right)  + U_{AB} \sum_{l} \hat a^\dag_{l}  \hat a_{l} \hat  b^\dag_{l}  \hat b_{l} \nonumber \\
  &+ \frac{U_A}{2} \sum_{l} \hat a^\dag_{l} \hat a^\dag_{l}  \hat a_{l} \hat a_{l}  + \frac{U_B}{2} \sum_l \hat b^\dag_{l} \hat b^\dag_{l} \hat  b_{l} \hat b_{l} 
 ,  \label{eq:two_spec_ham}
\end{align}
with $\hat a_l, \hat b_l$ bosonic annihilation operators for species $A$ and $B$ respectively, and where $\sum_{\langle j,l\rangle}$ denotes a sum over neighbouring sites. The adjustable microscopic separation between spin-up and spin-down sites is expressed as a tunable inter-component on-site energy $U_{AB}$, whereas the tunnelling amplitude for each species is $J$ and the intra-component interactions are $U_A$ and $U_B$. 

In the regime of large intra-species interaction $U_A=U_B\equiv U \gg J$, the two-species Mott Insulator with two atoms per site can be described by a pseudo-spin triplet, as depicted in Fig.~1a. In the case of unit filling with $N_A=N_B=L$ atoms and sites, model \eqref{eq:two_spec_ham} can be
mapped on a effective spin $S=1$ model in second order perturbation
theory \cite{Altman2003}. The effective states of spin in the $z$ direction are proportional to $a^\dag a^\dag
\ket{0}$ ($S^z=+1$), $a^\dag b^\dag\ket{0}$ ($S^z=0$), and $b^\dag b^\dag\ket{0}$ ($S^z=-1$), as shown in Fig.~1a. The effective model is
a ferromagnetic Heisenberg lattice or chain with Hamiltonian 
\begin{align}
 \label{eq:eff_ham_2}
  \mathcal{H}_{\rm eff} 
  =
 -J_{xyz} \sum_{\langle j, l\rangle} {\bf \hat S}_j {\bf \hat S}_{l}
 &+ u \sum _{l} (\hat S^z_l)^2,
\end{align}
where  $u=U-U_{AB}$, $J_{xyz}=4J^2/U_{AB}$, and we define ${\bf
  \hat S}_l= \left(\hat S_l^x, \hat S_l^y, \hat S_l^z \right)$.

As shown in Fig.~1b, the magnetic state depends on the interactions: for small inter-component repulsion, the ground state is the $S=1, S^z=0$ state, whereas for inter-component repulsion comparable to intra-component interactions, the ground state is an $xy$-ferromagnet, where each site is in a superposition of the $S^z= +1, 0, -1$ states \cite{Altman2003}.  The latter state features superfluid spin-transport (or counterflow superfluidity) \cite{Kuklov2003}, whereas the former is a spin insulator or spin Mott state.  By varying the relative positions of the spin-dependent lattices, we tune $U_{AB}$, as shown in Fig.~1c, adiabatically connecting the spin Mott state to the $xy$-ferromagnetic state \footnote{Note that for Rb atoms, since all scattering lengths are almost equal, $U_{AB}/U$ can be varied in a range between 0 and 1.}. We thus realize a quantum phase transition from a gapped state without any broken symmetries to a state which is magnetically ordered via superexchange. This is a superfluid to insulator transition in the spin domain.  For adiabatic cooling, the spin Mott state shares many advantageous features with the fermionic band insulator: they are both gapped, and the spins are already fully mixed, and only microscopic transport can connect the gapped phase to magnetically ordered phases.

\textit{Validation --} In the remainder of this paper, we validate this idea with t-DMRG calculations. We calculate ground-states and time-evolution in the full two-species model \eqref{eq:two_spec_ham}, truncating the total number of particles allowed on one site in the numerics to the value $n_{\rm max}$ \footnote{Note that the quantitative variation from the full bosonic model is very small, as discussed in the supplementary material \cite{SupMat}.},  and calculate spin-observables in the low-energy spin-subspace.  Spin-dependent lattices require near resonant laser light (detuned by less than the fine-structure splitting), which causes heating by spontaneous light scattering. Therefore, very slow adiabatic ramps are not possible, but as we show in this paper, there are parameter regimes where we can access the magnetically ordered phase. Although the many-body state fidelity is low, magnetic correlations still persist.  Since the Mott phase in 1D forms at much faster tunneling rates ($U/J\approx 3.3$) than in 3D ($U/J\approx 30$), we choose a 1D system to allow for faster ramps. The novel feature of our calculations is the combination of exact solutions for adiabatic ramps with a master equation for spontaneous emission of photons.  Technical and other noise can also easily be added.  In this sense, our study is a major step towards fully realistic simulations of experimental schemes for accessing new quantum phases.

\begin{figure}[t]
  \centering
\includegraphics[width=\columnwidth]{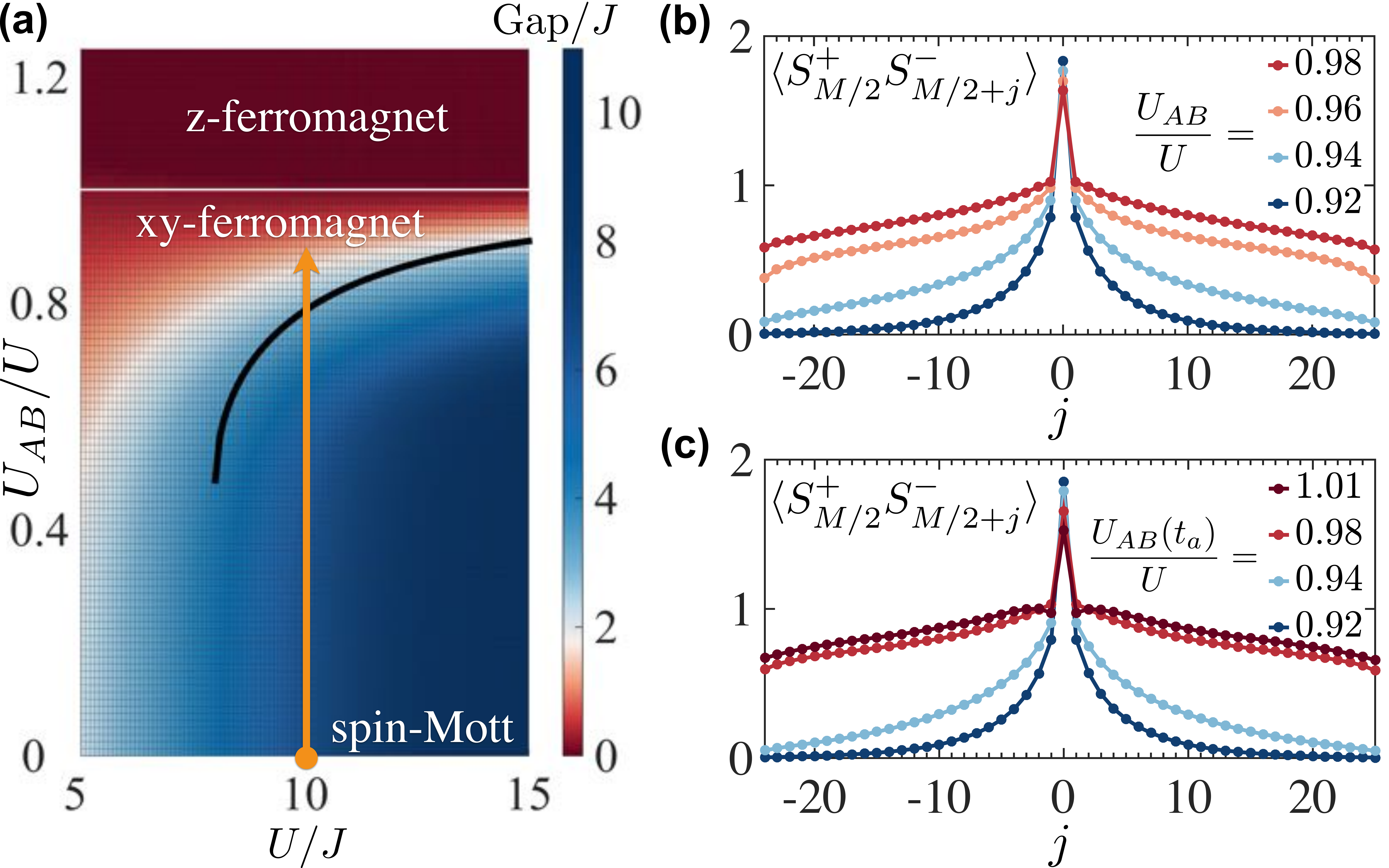}
\caption{{\it Magnetic phase diagram and correlations --} (a) Phase-diagram for two-component bosons in a 1D optical lattice. The color coding shows the gap in a small system with 12 particles on 6 lattice sites. The black solid line indicates the mean-field phase transition from a spin Mott to an $xy$-ferromagnetic phase. Our adiabatic ramp is along the orange arrow. Along this path a phase transition to the $xy$-ferromagnet occurs at $U_{AB}/U = 0.942 \pm 0.001$ \cite{SupMat}.  (b) The $xy$-ferromagnetic ground-state is characterized by the onset of algebraically decaying $\langle S^+_i S^-_{i+j} \rangle$ correlations (DMRG calculations for 100 bosons on 50 lattice sites, $U=10J$, $n_{\rm max}=4$) (c) The same type of correlations as in (b) but now obtained with a time-dependent ramp with a final ramp speed of $dU_{AB}/dt=0.01J^2$ (t-DMRG calculation, $n_{\rm max}=4$). } \label{fig:2}
\end{figure}

\textit{Phase diagram and spin correlations --} In Fig.~2a, we show a sketch of the phase diagram of model \eqref{eq:two_spec_ham}. A mean-field calculation shows that the phase transition in the spin picture occurs at $u/J_{xyz}=4$ \cite{Solyom1984}, or $U_{AB}/U = 1/2 + (\sqrt{1-64/(U/J)^2})/2$, shown as a thick black line in the figure. In 1D making use of a DMRG calculation \cite{SupMat, Deng2005}, we find a surprisingly large shift of the phase transition from the mean field value, e.g., from $U_{AB}/U=0.8$ to $U_{AB}/U = 0.942 \pm 0.001$ for $U/J=10$ \footnote{Note that the quoted phase transition point is estimated for $n_{\rm max}=4$ \cite{SupMat}.}. The shading in the figure represents the energy gap between the ground and lowest excited states in a system with 12 particles on 6 lattice sites. This indicates where an adiabatic ramp will be most difficult in a finite-size system.

To identify the $xy$-ferromagnetic ground-state, we study spin-spin correlation functions of the form $\langle S^+_l S^-_{l+j} \rangle$. Outside the $xy$-ferromagnetic regime, these correlations decay exponentially, whereas they decay algebraically in 1D on the $xy$-ferromagnetic side of the transition. In Fig.~2b, we see clearly the qualitative change in behaviour across the transition in the ground-state spin-spin correlation functions, which could be detected via noise correlation imaging \cite{Altman2003}.

\textit{Calculation of adiabatic ramps --} We now validate the ramp procedure for finite-size systems of the scale that will typically be present in cold atom experiments. Beginning in a spin Mott state with $U_{AB}\approx 0$, we initially increase $U_{AB}$ rapidly at a constant rate of $dU_{AB}/dt=1J^2$ to a value of
$U_{AB}/U=0.75$. This rapid ramp is adiabatic because of the large spectral gap. We then use a second, slower ramp to the final state, again at a constant rate. Note that such ramps could be significantly further optimized by quantum control techniques, making the estimates for timescales given here very conservative. The correlation functions at the end time of the ramp, $t_a$ are shown for different values of $U_{AB}$ in Fig.~2c, and are almost identical to those in the ground-state up to $U_{AB}=0.98U$. 

For $U_{AB}>U$, the ground state is a $z-$ferromagnet, which for a constant number of particles amounts to phase separation of the atoms. However, the symmetry change between the $xy$-ferromagnet and the $z-$ferromagnet prevents this transition from occurring adiabatically. We find that if we ramp across the transition, instead we produce a metastable excited state in which the $xy$-ferromagnetic correlations persist, as shown in Fig.~2c for $U_{AB}=1.01U$. 

As a stringent test of adiabaticity, we calculate the fidelity of the quantum state throughout the ramp, defined as
\begin{align}
  \label{eq:1}
  \mathcal{F}= | \langle\psi_{\rm gs}(U_{ab})|{\psi(t)}\rangle|^2
  ,
\end{align}
where $\ket{\psi(t)}$ denotes the time-evolved state during the ramp, and
$\ket{\psi_{\rm gs}(U_{ab})}$ the corresponding ground-state. We plot this in Fig.~3a as a function of $U_{AB}$, for different $t_a$. We see that for all ramps, the fidelity is very high until near the transition point, and for faster ramps falls rapidly at the transition to the $xy$-ferromagnetic regime. However, for long ramps, the state fidelity can approach $\mathcal F=1$. 

A key question in this context is how the timescale required for an adiabatic ramp depends on system size. We expect that for large systems, complete adiabaticity will be impossible as the gap to excited states goes to zero, and correlations will only be established over length scales shorter than the system size. However, as shown in Fig.~3b, it is possible for typical experimental system sizes to reach almost unit fidelity for ramps of realistic durations. For system sizes up to 50 lattice sites, a high-fidelity final state can be produced with ramps that are less than a second in duration.

\begin{figure}[tb]
  \centering
\includegraphics[width=\columnwidth]{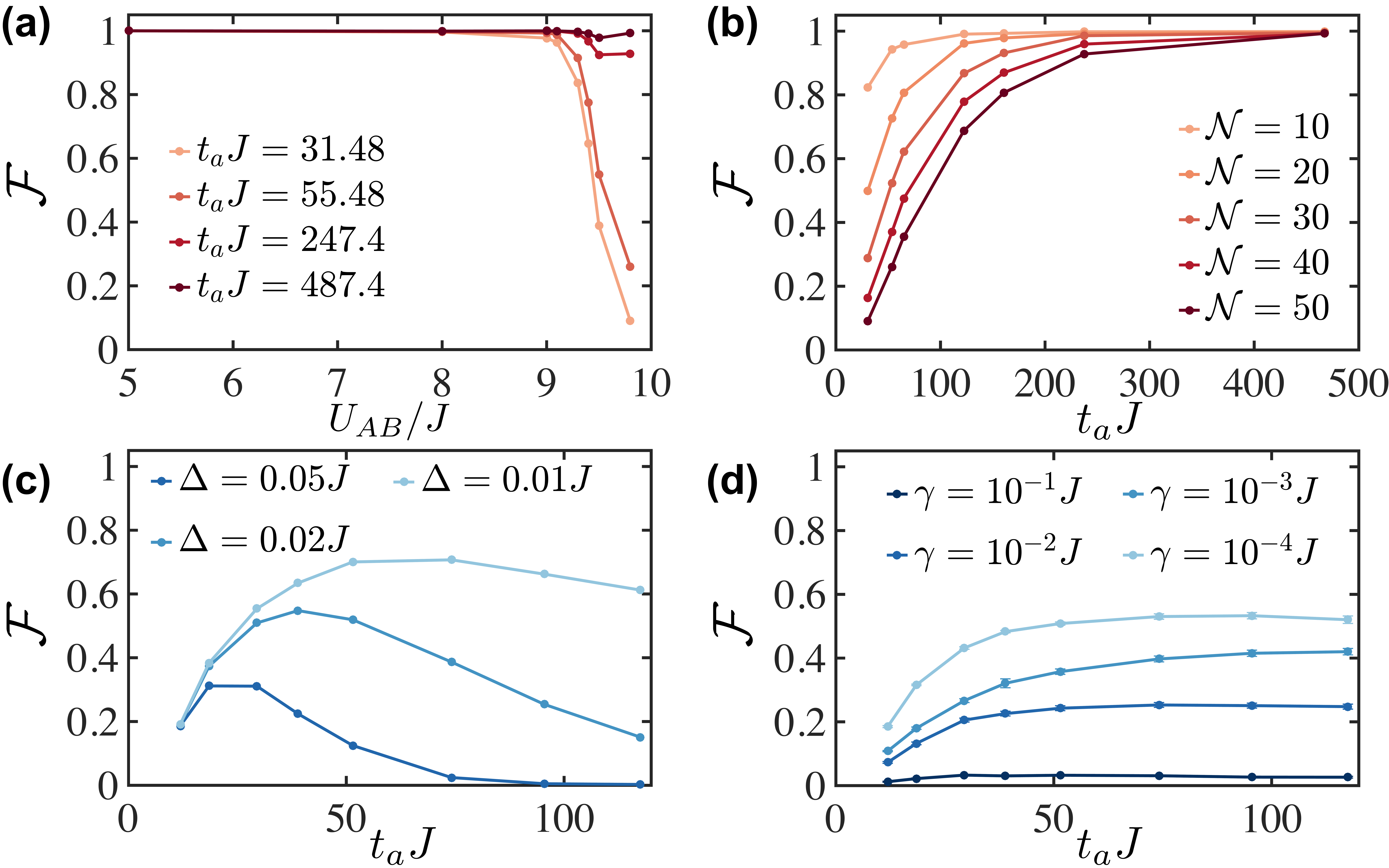}
\caption{{\it Many-body state fidelity $\mathcal{F}$ during adiabatic ramps --} (a) The fidelity of the adiabatically evolved state for different ramp times in a system with $100$ particles on $50$ sites. The fidelity reduces when crossing the phase-transition point at $U_{AB}/U\sim 0.94$. For slower ramps, a larger fidelity can be achieved. (b) The fidelity with which the $xy$-ferromagnetic state at $U_{AB}/U=0.98$ can be prepared as function of the preparation time and for different system sizes $\mathcal{N}$. With increasing $\mathcal{N}$, a larger preparation time is required to reach high state fidelities. (c/d) State fidelities for the $U_{AB}/U=0.97$ state in a system of $40$ particles with competing processes. (c) $\mathcal{F}$ for different magnetic field gradients $\Delta$. (d) $\mathcal{F}$ in the presence of spontaneous emissions with rates $\gamma$. The quantum noise dramatically reduces the achievable state fidelities and there is an optimum speed for the ramp [(a-c) $n_{\rm max}=4$, (d) $n_{\rm max}=3$].  \label{fig:3}}
\end{figure}

\begin{figure}[tb]
  \centering
\includegraphics[width=\columnwidth]{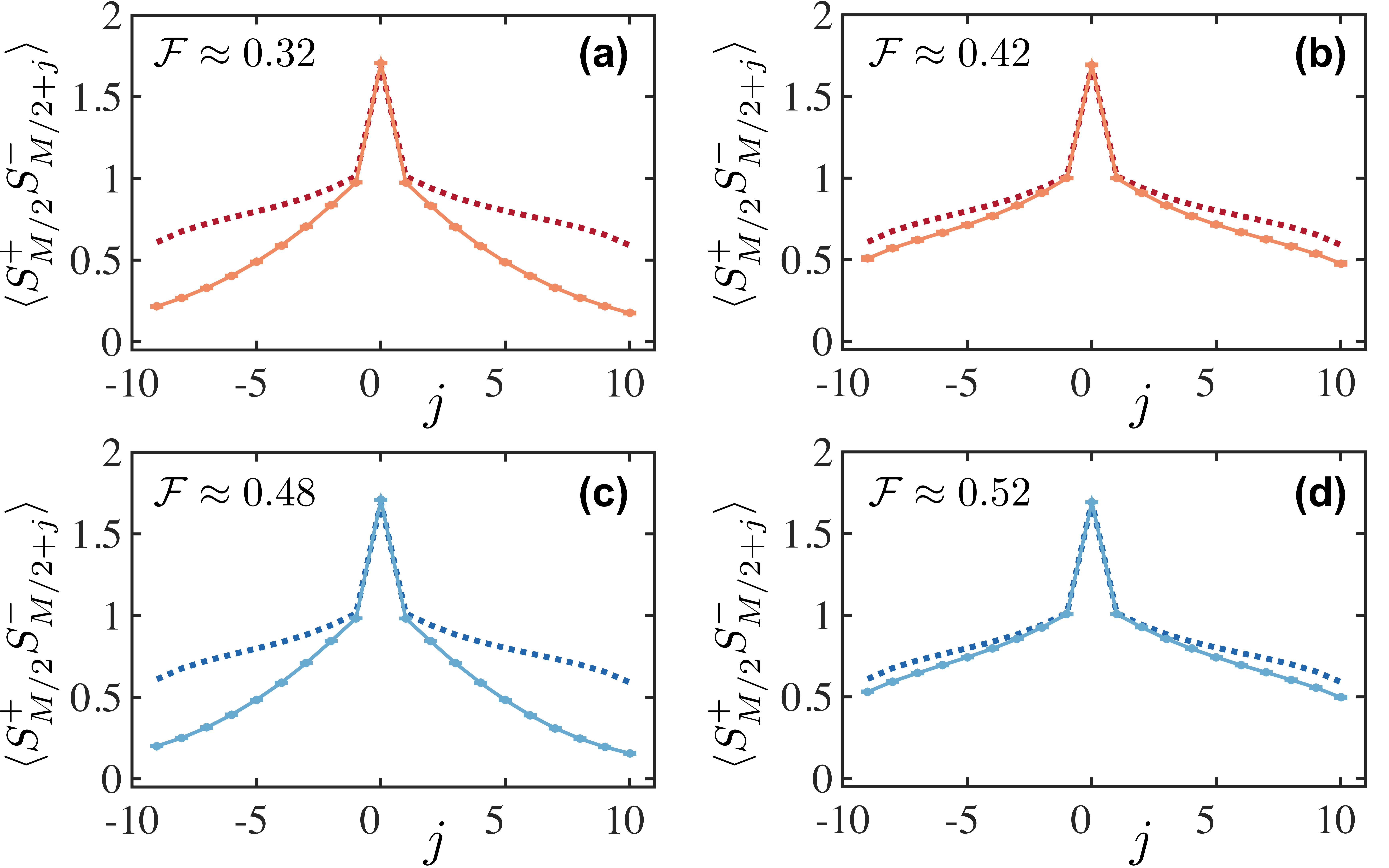}
\caption{{\it Comparison of correlations and state fidelity in the presence of spontaneous emission--}  $\langle S^+_i S^-_{i+j} \rangle$ correlations for the adiabatically prepared state with $U_{AB}/U=0.97$ in a system with $40$ particles on $20$ sites, compared to the ground-state correlation (dashed lines). (a/b): High spontaneous emission rate, $\gamma=10^{-3}J$, (c/d) low spontaneous emission rate $\gamma=10^{-4}J$. Panels (a) and (c) are for a faster ramp with $t_aJ=38.9$, (b) and (d) for the slower ramp with $ t_aJ=117.5$.  The state-fidelities are given in the plots.  \label{fig:4}}
\end{figure}

\textit{Competition from decoherence via spontaneous emissions --} The natural question is how these ramps compete with natural heating processes in the experiment. This leads to a trade-off between using faster ramps to avoid additional heating, and slower ramps to improve adiabaticity. An example of this competition is shown in Fig.~3c, where we show the final state fidelity if we consider the original ramp and ground-states of \eqref{eq:two_spec_ham}, but include a magnetic gradient potential term $\Delta \sum_l \, l a^\dag_l a_l$ in calculating the dynamics. As $\Delta$ is increased, the optimal ramps become shorter and achieve lower total fidelity, as the state is rotated away from the original model. Note that because the spin Mott state is robust against this potential, the main influence of this term comes only at the end of the ramp, reducing adiabaticity and dephasing the $xy$-ferromagnetic ordering. 

For spin-dependent lattices, the dominant heating mechanism will be spontaneous emissions at an effective scattering rate $\gamma$. For a typical setup with Rubidium atoms, the dynamics will then be dominated by localisation of particles that remain in the lowest band of the lattice \cite{Gerbier2010,Pichler2010}, which can be described microscopically by a master equation for the system density operator $\rho$ \cite{Pichler2010},
\begin{equation}
\label{eq:QT_master_eq}
\dot{\rho}=-\frac{\rm i}{\hbar} [H,\rho]-\frac{\gamma}{2}\sum_{i}\left( n_i n_i\rho+\rho n_i n_i-2n_i\rho n_i \right),
\end{equation}
with $n_i=a^\dag_i a_i + b^\dag_i b_i$. We solve this master equation by combining t-DMRG methods with quantum trajectories techniques \cite{Daley2014} to obtain a complete microscopic description including heating. In Fig.~3d we then plot the fidelity as a function of $t_a J$ for different $\gamma$ values. Again, we see a trade-off between heating and adiabaticity, leading to very low maximal fidelities for large heating rates.

While in the absence of heating, fidelities characterize the adiabaticity and thus also the quality of the final magnetic correlations relatively well, this is not the case in the presence of heating. In fact, the magnetic correlations exhibit a surprising degree of robustness against heating due to spontaneous emissions. In Fig.~4 we plot correlation functions at the end of the ramps in the presence of spontaneous emissions. Especially by comparing the lower fidelity state in Fig.~4b and the higher fidelity state in Fig.~4c, we see that the strength of correlations is disconnected from the fidelity. It is actually advantageous to use longer ramps despite a reasonable increase in spontaneous emissions, and as demonstrated in Fig.~4d, strong magnetic correlations are achievable for typical system sizes after scattering of the order of 5 photons, despite the large energy that would be introduced in comparison with the superexchange energy $J^2/U$.   

\textit{Outlook --} We have demonstrated that the spin Mott state of two-component bosons can be used as a starting point for producing sensitive, ordered many-body states via adiabatic ramps, and at the same time that the combination of t-DMRG and quantum trajectories can be used to fully address possible experimental limitations, and provide a microscopic guide to adiabatic state preparation. These experimental and theoretical techniques can be immediately generalised to produce a rich array of many-body states, including regimes accessible in mass-imbalanced bosonic or Bose-Fermi mixtures.

\textit{Acknowledgements --}
We thank Colin Kennedy, Stephan Langer, Hannes Pichler, and Saubhik Sarkar for stimulating discussions. This work was supported in part by the EOARD via AFOSR grant number FA2386-14-1-5003, by AFOSR MURI FA9550-14-1-0035, by ARO-MURI Grant No. W911NF-14-1-0003, by the NSF through grant PHY-0969731, through JILA under grants JILA-NSF-PFC-1125844, NSF-PIF-1211914, through the Center for Ultracold Atoms, and by the Aspen Center for Physics with support under NSF grant 1066293. Computations utilized the Janus supercomputer, supported by NSF (CNS-0821794), NCAR, and CU Boulder/Denver, and code development was supported by AFOSR grant FA9550-12-1-0057.

   \def\eprint#1{arXiv:#1}

\bibliographystyle{apsrev}
\bibliography{adiabaticbosons}

\pagebreak

\onecolumngrid
\appendix
\section{Supplementary material: Determination of the ground-state phase diagram}
Here we present details on our determination of the ground-state phase-transition point via DMRG calculations and compare results from the full two-species model \eqref{eq:two_spec_ham} and the effective $S=1$ spin-model \eqref{eq:eff_ham_2}. Because we are in the balanced (equal particle number) regime $N_A=N_B$, the magnetization itself is always zero. Instead, we can determine the transition point from fluctuations in the planar magnetization, $\delta_{xy}\equiv \sum_l \langle (\hat S^x_l)^2 \rangle + \langle (\hat S^y_l)^2 \rangle$, which we expect to have a discontinuity in its second derivative with respect to $U_{AB}$ in the thermodynamic limit, where this quantity jumps between a positive value and a negative value. For a finite system, this becomes a zero crossing, and we determine the point at which this occurs numerically, in analogy e.g., to Ref.~\cite{Deng2005}. We note that this quantity should be experimentally accessible since it is equivalent to the number of doubly occupied sites, $(\hat S^x_l)^2 + (\hat S^y_l)^2= \hat a^\dag_l  \hat b^\dag_l \ket{0}\bra{0} \hat b_l \hat a_l + \mathbb{1}$.

\begin{figure}[h]
  \centering
\includegraphics[width=0.8\columnwidth]{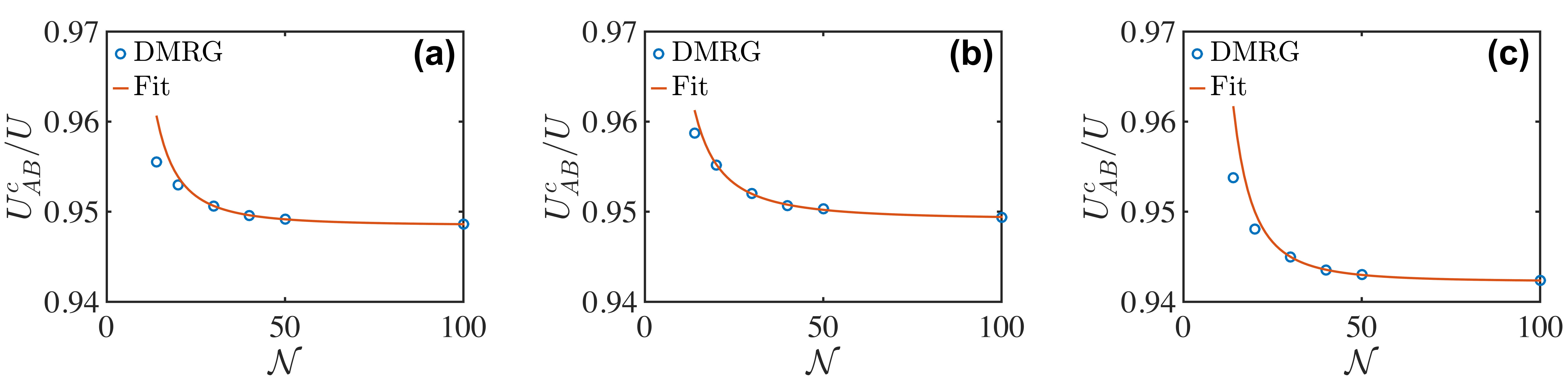}
\caption{{\it Ground-state phase transition determination--}  DMRG phase-transition estimations in systems with various sizes $\mathcal{N}$ as well as a finite-size scaling fit for: (a) The effective $S=1$ spin model \eqref{eq:eff_ham_2}, (b) a two-species model  \eqref{eq:two_spec_ham} with $n_{\rm max}=3$; and (c) the two-species model with $n_{\rm max}=4$. In each case, the transition point is determined by finding the value of $U_{AB}/U$ at which a sign change occurs in the second derivative of the total fluctuations of the planar magnetization, $d^2\delta_{xy}/dU_{AB}^2$. \label{fig:A1}}
\end{figure}

To determine the transition point we use DMRG techniques to calculate $\delta_{xy}$ for $U=10J$ and multiple values of $U_{AB}$ with increments of $0.01J$. Using a spline interpolation we then numerically estimate the zero-crossing of $d^2\delta_{xy}/dU_{AB}^2$. We repeat this procedure for different system sizes of $\mathcal{N}=14,20,30,40,50,100$ sites. Fig.~\ref{fig:A1} summarizes the results of the estimated critical value of $U_{AB}^c$ as function of $\mathcal{N}$ from our procedure for the different models.

We find that for $\mathcal{N}\geq 30$, with small error $U_{AB}^c(\mathcal{N})$ follows a power law of the form $U_{AB}^c(\mathcal{N})=U_{AB}^c(\mathcal{\infty})+{\rm const.} \times \mathcal{N}^{-\alpha}$. We use a fit to obtain estimates for $U_{AB}^c(\mathcal{\infty})$ with a statistical fitting error to our finite size scaling. To summarize, for the different models we find: i) S=1 spin model, $U_{AB}^c(\mathcal{\infty})=9.48\pm0.01J$; ii) two-species model with $n_{\rm max}=3$, $U_{AB}^c(\mathcal{\infty})=9.49\pm0.01J$; and iii) two-species model with $n_{\rm max}=4$, $U_{AB}^c(\mathcal{\infty})=9.42\pm0.01J$. 

The two-species model with $n_{\rm max}=3$ is closer to a hard-core model that is restricted to the effective spin-basis and thus the estimated transition point is closer (identical within errors) to the one of the spin-model in this case. We find a small quantitative variation of $U_{AB}^c(\mathcal{N})$ for $n_{\rm max}=4$, but the qualitative behaviour of the dynamics in the effective spin sub-space is unaffected by variations of $n_{\rm max}$. 

In our main manuscript we use $n_{\rm max}=4$ except for the open system calculations, where to enable greater numerical accuracy and efficiency, we restrict to a basis with $n_{\rm max}=3$. Based on the small quantitative shifts we observe for the phase transition point, and our understanding of the dynamics obtained from the $n_{\rm max}=3$ case, we expect that the quantitative variations from the full model should be comparable to our statistical errors. 

\end{document}